\documentclass[aps,twocolumn,pra,tightenlines,floatfix,showpacs]{revtex4}

\usepackage[dvips]{graphicx}
\usepackage[english]{babel}
\usepackage{amsmath}
\usepackage{amssymb}
\usepackage{times}

\newcommand{\bs}{\begin{split}}
\newcommand{\es}{\end{split}}
\newcommand{\be}{\begin{equation}}
\newcommand{\ee}{\end{equation}}
\newcommand{\ba}{\begin{eqnarray}}
\newcommand{\ea}{\end{eqnarray}}

\begin{document}

\title{Superfuid-insulator transitions at non-integer filling in
  optical lattices of fermionic atoms}

\author{Chih-Chun Chien, Yan He, Qijin Chen, and K.  Levin}

\affiliation{James Franck Institute and Department of Physics,
 University of Chicago, Chicago, Illinois 60637}

\date{\today}

\begin{abstract}
  We determine the superfluid transition temperatures $T_c$ and the
  ground states of the attractive Hubbard model and find new insulating
  phases associated with non-integer filling at sufficiently strong
  pairing attraction $|U|$. These states, distinct from band and
  Mott insulating phases, derive from pair localization; pair hopping at
  large $|U|$ and high densities is impeded by inter-site, inter-pair
  repulsive interactions.  The best way to detect the breakdown of
  superfluidity is using fermionic optical lattices which should reveal
  new forms of ``bosonic'' order, reflecting ground state pairing
  without condensation.

\end{abstract}


\maketitle

%% ---------------------------------------------------------------------- 
The attractive Hubbard Model has captured the attention \cite{MicnasRMP}
of the theoretical and experimental communities, although this model is
far from fully understood. It is hoped that experiments using ultracold
fermionic atoms in optical lattices \cite{Ketterlelattice} will shed
light both on competing ground states and on finite temperature $T$
effects, and moreover, provide insights into high temperature
superconductivity, largely through the very ubiquitous ``pseudogap''
effects. These derive from \cite{Reviews} the fact that as the
attraction becomes stronger, pairing takes place at a different
temperature, $T^*$ from superfluid condensation, $T_c$.

For bosonic systems at half filling the strong (repulsive) interaction
limit gives rise to superfluid-Mott insulator transitions
\cite{oldBloch}.  It is a central goal of this paper to investigate this
counterpart strong (attractive) interaction regime for fermions and
demonstrate that superfluid-insulator (SI) transitions (away from
integer filling) are also present.  Indeed, the latest experiments on
fermions\cite{Ketterlelattice} reveal competing insulating and
superfluid ground states. It is not yet clear whether these insulating
phases are band insulators or localization-induced insulating phases.
We note that the largest coupling regime we address in this paper has
not yet studied by Monte Carlo \cite{Beck02} or other methods
\cite{DMFT}, despite the fact that a comparably strong (dimensionless)
interaction strength is required for bosonic systems to undergo SI
transitions.

Our finite $T$ approach to the Hubbard model includes pair fluctuations
in a manner which is consistent with the standard BCS-Leggett ground
state and, most importantly, with a proper and physical treatment of
pseudogap effects.  We stress, however, that these non-condensed pair
effects must necessarily be treated differently from previous
Nozi\`eres--Schmitt-Rink \cite{NSR} based approaches to $T \neq 0$.
Indeed, following Ref.~\cite{NSR}, essentially all other pairing
fluctuation approaches to BCS-BEC crossover contain an inherent
inconsistency; they presume that in the fermionic dispersion relation
$E_{\bf k} = \sqrt{ \xi_{\bf k}^2 + \Delta^2 (T) }$, the so-called
pairing gap $\Delta$ vanishes at and above $T_c$.  These non-condensed
pairs (present for all $0< T < T^*$) are also essential for arriving at
physical values for the transition temperature $T_c$.  Without including
them it is not possible to know about ground states with $T_c=0$, which
will naturally occur, \textit{e.g.}, in the present theory \cite{Chen1}
in two-dimensional lattices, as compatible with the Mermin-Wagner
theorem (in the absence of Kosterlitz-Thouless or other topological
order).  This underlines the fact that one cannot solve the BCS-Leggett
$T=0$ equations in isolation to characterize the stable ground states.

Our principal result is that, in addition to the expected band
insulating state (at filling $n=2$), a new insulating phase, which is
stabilized by pair fluctuations, emerges when $n$ and the interaction
strength exceed appropriate critical values.  Strikingly, the critical
value for the dimensionless interaction strength is comparable to that
found for the Bose Hubbard model \cite{oldBloch}. This new insulating
phase is different from the Mott or band insulator, both of which occur
only at integer filling, but like the Mott case it is due to
localization (of fermion pairs).  To address cold atom optical lattices,
we also extend the attractive Hubbard model to a two-channel model and
find that the insulating phases at non-integer filling survive; thus,
pair localization is a robust effect.

Our work can be contrasted with previous studies, which focused only on
$T=0$ \cite{Duan05,Carr05,Zhou05,Holattice} and with Monte Carlo (MC)
\cite{Beck02} as well as dynamical mean field theoretic approaches
\cite{DMFT}.  In the moderate coupling regime, where comparisons can be
made, we find transition temperatures to be of comparable order of
magnitude, albeit slightly larger (factors of 2).  In part this derives
from the fact that 
the standard BCS-Leggett
ground state considered here does not include \cite{DMFT} the 
Gorkov--Melik-Barkhudarov
\cite{HeiselbergGMB} effect.  This semi-quantitative
difference should not be of concern
because different MC calculations \cite{Amherst,BulgacPRL} of $T_c$ in a
unitary Fermi gas have not yet converged to better than factors of 1.5.

Of particular interest here are the properties of the attractive Hubbard
model, in the limit that the attraction $|U|$ is much larger than the
hopping $t$.  While this model exhibits a smooth crossover from BCS to
Bose-Einstein condensation (BEC) with increasing $|U|$, it should be
stressed that \textit{the BEC here is very different from that of true
  bosons}, since the hopping of the pairs involves, as an intermediate
or virtual step, their unbinding into fermions. In this way the inverse
pair mass was argued \cite{NSR,MicnasRMP} to vary as $t^2/|U|$. Important
for the present purposes is the fact that \textit{there is also a
  nearest neighbor repulsion \cite{NSR,MicnasRMP} which is of the same
  magnitude as the effective pair hopping term}.  On this basis it has
been argued
% one can anticipate, then, a frustration of the motion of the pairs.
% Indeed, Micnas has argued
that \cite{MicnasRMP} ``at higher densities the overlap of the bound
pairs severely restricts their motion''. This quotation is particularly
relevant to our findings; we show that at sufficiently high $n$ the pair
motion is so restricted that their effective mass diverges prematurely
before $|U| \rightarrow \infty$.  This localization, in turn, is
associated with the breakdown of superfluidity.

This frustration of superfluidity at high $n$ can also be addressed via
a mapping to a magnetic model.  When $|U|/t\gg 1$ the attractive Hubbard
model is equivalent to \cite{MicnasRMP} an effective quantum XXZ model
with coupling constants $J_{XX}=-J_{Z}$, both of which are proportional
to $t^{2}/|U|$.  This mapped problem is subject to the constraint that
the average magnetization along the $z$-axis is fixed at $(n-1)/2$,
making this a rather difficult magnetic model to solve in general.
Superfluidity corresponds to ordering in the $x-y$ plane.  When $n$ is
small, the constraint is straightforward to implement and superfluidity
emerges.  However, at high densities ($n \approx 1$) new states with
order along the $z$-axis are expected \cite{MicnasRMP} to emerge,
thereby, destroying the superfluid phases.
Indeed, it is well known that for $n$ strictly equal to 1, the
simple superfluid ground state is undermined as a result of a degeneracy
with a charge ordered ground state \cite{MicnasRMP}.

We now address these effects by investigating the physics of fermions
which interact via $s$-wave attraction in a three dimensional (3D)
square lattice, we start with the one-channel attractive Hubbard model
Hamiltonian
\begin{eqnarray}
H_{f}&=&\sum_{\mathbf{k}\sigma}\xi_{\mathbf{k}}c^{\dagger}_{\mathbf{k}\sigma}c_{\mathbf{k}\sigma} \nonumber \\
&&{}+ U\sum_{\mathbf{kk^{\prime}q}}
c^{\dagger}_{\mathbf{k+q}/2\uparrow}c^{\dagger}_{-\mathbf{k+q}/2\downarrow}c_{-\mathbf{k^{\prime}+q}/2\downarrow}c_{\mathbf{k^{\prime}+q}/2\uparrow}.
\end{eqnarray} 
Here $\xi_{\mathbf{k}}=\epsilon_{\mathbf{k}}-\mu$, where $\mu$ is the
chemical potential. The one-particle energy dispersion is
$\epsilon_{\mathbf{k}}=2t[3-\cos k_x-\cos k_y -\cos k_z]$ in a one-band
nearest-neighbor tight-binding approximation, where the values of
$\mathbf{k}$ are restricted in the first Brillouin zone, and we set
lattice constant $a_0=1$.  $U$ represents the attractive on-site
coupling.  ``Resonant'' scattering, which corresponds to an infinite two
body scattering length, occurs at $U/t\approx -7.915$, in agreement with
Ref.~\cite{Troyer06}.  Note that by adopting the Hubbard model we drop
any terms associated with direct pair hopping which should not be
important in the regime we focus on here where $\mu$ is positive or only
slightly negative.  We consider a one band model, on the premise that
multi-band effects will change the results quantitatively but not
qualitatively.  There is still uncertainty in the literature
\cite{Holattice} about whether or not an effective one band
\cite{Duan05} model is adequate.

We use a $T$-matrix formalism \cite{Reviews,Chen4,JS2} to address finite
temperature.  This particular $T$-matrix approach has a crucial
advantage because it leads to physical results for the superfluid
density $n_s(T)$.  This single valued, monotonic and continuous behavior
(from zero to $T_c$) for $n_s(T)$ is not found in other theories; this
physical behavior can be traced to a self consistent treatment of
pseudogap effects \cite{Reviews}, in which pair fluctuations enter into
\textit{both} the gap and the number equations in a fully
self-consistent fashion

Details of this formalism can be found in \cite{Reviews}. We define the
noncondensed pair propagator as $t_{pg}(Q)= U/[1+U\chi(Q)]$, where, as
in Ref.~\cite{Chen2}, our choice for the pair susceptibility, given by
$\chi(Q) = \sum_{K}G_{0}(Q-K)G(K)$, can be derived from decoupling the
Green's function equations of motion.  Here $G(K)$ and
$G_{0}(K)=i\omega_n -\xi_{\mathbf{k}}$ are the full and bare Green's
functions. $K\equiv (i\omega_n, \mathbf{k})$ and $Q\equiv (i\Omega_m,
\mathbf{q})$ are four-vectors with $\sum_{K}\equiv
T\sum_{\omega_n}\sum_{\mathbf{k}}$. Below $T_c$, the self-energy
$\Sigma(K)=\sum_{Q}t(Q) G_{0}(Q-K)$ can be well approximated by the BCS
form, $\Sigma(K) = -\Delta^2 G_{0}(-K)$, where
$t(Q)=-(\Delta_{sc}^2/T)\delta(Q)+t_{pg}(Q)$, and $\Delta_{sc}$ is the
superfluid order parameter. In the superfluid state, the ``gap
equation'' is given by the pairing instability condition
$t_{pg}^{-1}(0)=U^{-1}+\chi(0)=0$, which is equivalent to the BEC
condition on the pairs.  Therefore, we have
\begin{equation} \label{eq:1geq}
\frac{1}{U}=-\sum_{\mathbf{k}}\left[\frac{1-
    2f(E_{\mathbf{k}})}{E_{\mathbf{k}}} \right].  
\end{equation}
Here
$f(x)$ is the Fermi distribution function. Similarly, the average
\cite{Kohl05a} density $n$ in a lattice, derived from
$n=\sum_{K,\sigma}G_{\sigma}(K)$, is given by
\begin{equation} \label{eq:1neq} n = \sum_{\mathbf{k}}
  \left[\left(1-\frac{\xi_k}{E_k}\right) +
    2f(E_k)\left(\frac{\xi_k}{E_k}\right)\right].
\end{equation}

We next determine the dispersion relation and the number density for
\textit{noncondensed} pairs.  From the self-energy expression one
obtains $\Delta^2=\Delta_{sc}^{2}+\Delta_{pg}^{2}$ where the pseudogap
contribution satisfies \cite{Reviews,JS2,Chen2}
\begin{equation}
  \Delta_{pg}^{2}\equiv - \sum_{Q}t_{pg}(Q) \,, \label{eq:pgeq}
\end{equation} 
which can be shown to vanish at $T=0$.  The critical temperature $T_c$
is defined as the lowest temperature where $\Delta_{sc}=0$.  In the
superfluid phase (pair chemical potential $\mu_{pair}=0$), and at small
$Q$, $t_{pg}^{-1}(Q)= \chi (Q)-\chi(0)\approx
Z_{1}\Omega^{2}+|Z_{0}|\Omega- \xi^2 q^2$.  Except when particle-hole
symmetry is present, e.g., at very weak coupling and near half filling,
we find $Z_{1}\ll |Z_{0}|$, which thus is irrelevant.  Near $n=1$, the
$Z_{1}\Omega^{2}$ term enters and regularizes the van Hove singularity.
To first order in $\Omega$ one can write
$\Delta_{pg}^{2}=|Z_{0}|^{-1}\sum_{\mathbf{q}} b(\Omega_{\mathbf{q}})$,
where $b(x)$ is the Bose distribution function.  At sufficiently low
$\mathbf{q}$, we can approximate the dispersion of the noncondensed
pairs by $\Omega_q = q^2/2M^{*}$, where $M^{*}$ is the effective mass of
pairs at the bottom of the band.
 
\begin{figure} 
\centerline{\includegraphics[clip,width=3.4in]{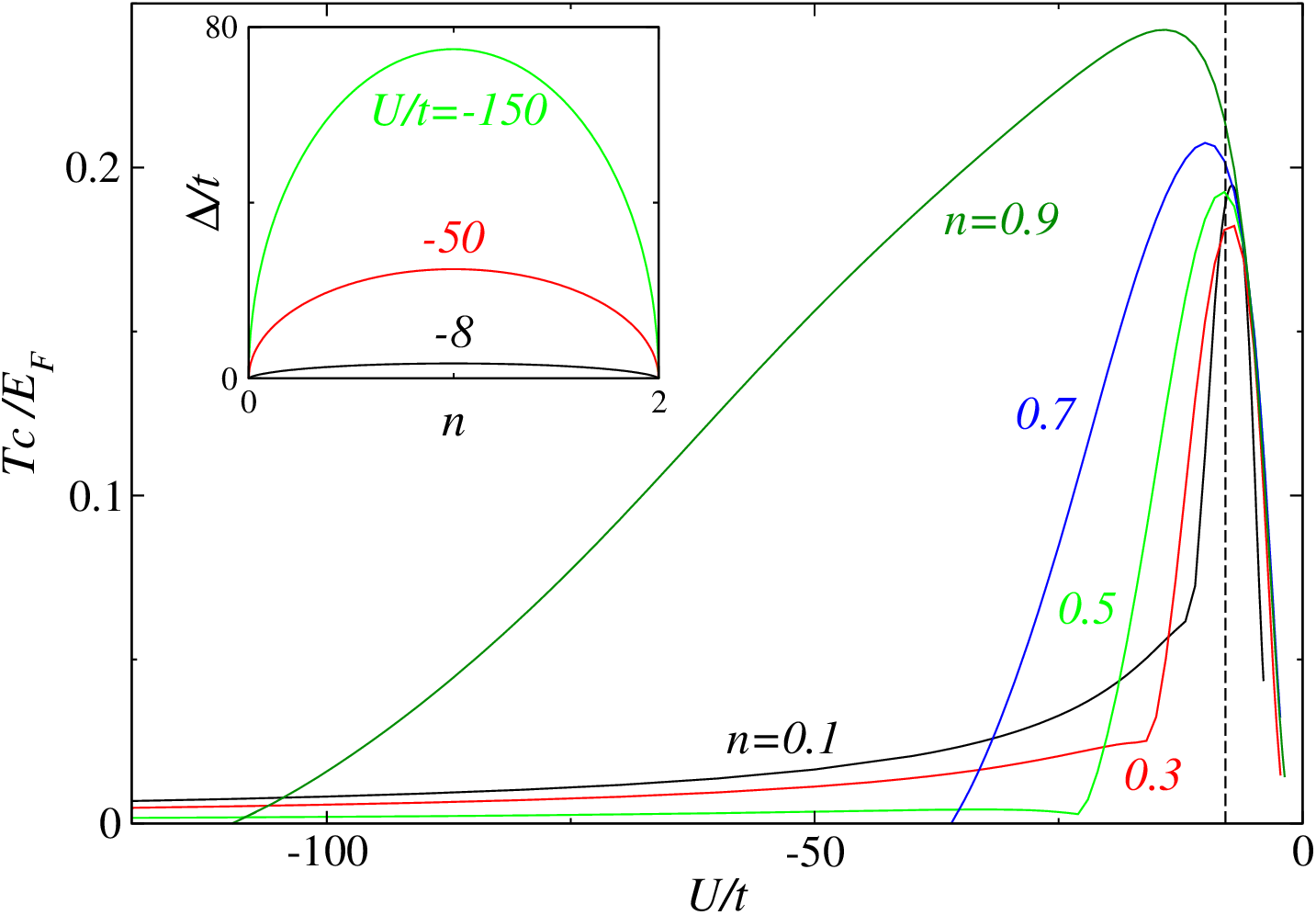}}
\caption{\label{fig:L1Tc} $T_c/E_{F}$ as a function of $U/t$. The
  corresponding value of $n$ is marked next to each curve. The vertical
  dashed line indicates the unitary limit. Inset: $T=0$ mean-field
  results for $\Delta /t$ as a function of $n$.}
\end{figure}

In Fig.~\ref{fig:L1Tc} we show $T_c/E_{F}$ as a function of $U/t$ for
several values of $n$ in the one-channel model. Here $E_{F}$ is the
Fermi energy of a non-interacting Fermi gas with the same filling
factor.  As in MC simulations, $T_c$ exhibits a maximum near resonance
and decreases slightly when $U/t$ is away from resonance. When $n$ is
small, we find a long tail in $T_c$ proportional to $t^2/|U|$, as
expected \cite{NSR}.  Importantly, it is observed that when
$n>n_c\approx 0.53$, $T_c$ vanishes provided $U/t$ exceeds a critical
value $(U/t)_c$.  It should be noted that at the insulating onset point,
we find $\mu$ is zero or slightly positive, far from the true bosonic
regime.
The inclusion of pair fluctuation effects is essential here for
establishing $T_c=0$ beyond $(U/t)_c$. This result can be contrasted
with the predicted ground state of strict mean-field theory shown in the
inset of Fig.\ref{fig:L1Tc}, where $\Delta /t$ vanishes only at $n=0$
and $n=2$ (band insulator).
%Although the maximum of $T_c$ close to resonance is consistent with
%Monte Carlo simulations \cite{Beck02}, the range of $U/t$ which has been
%studied in these simulations is limited to the vicinity of unitarity.

\begin{figure} 
\centerline{\includegraphics[clip,width=3.4in]{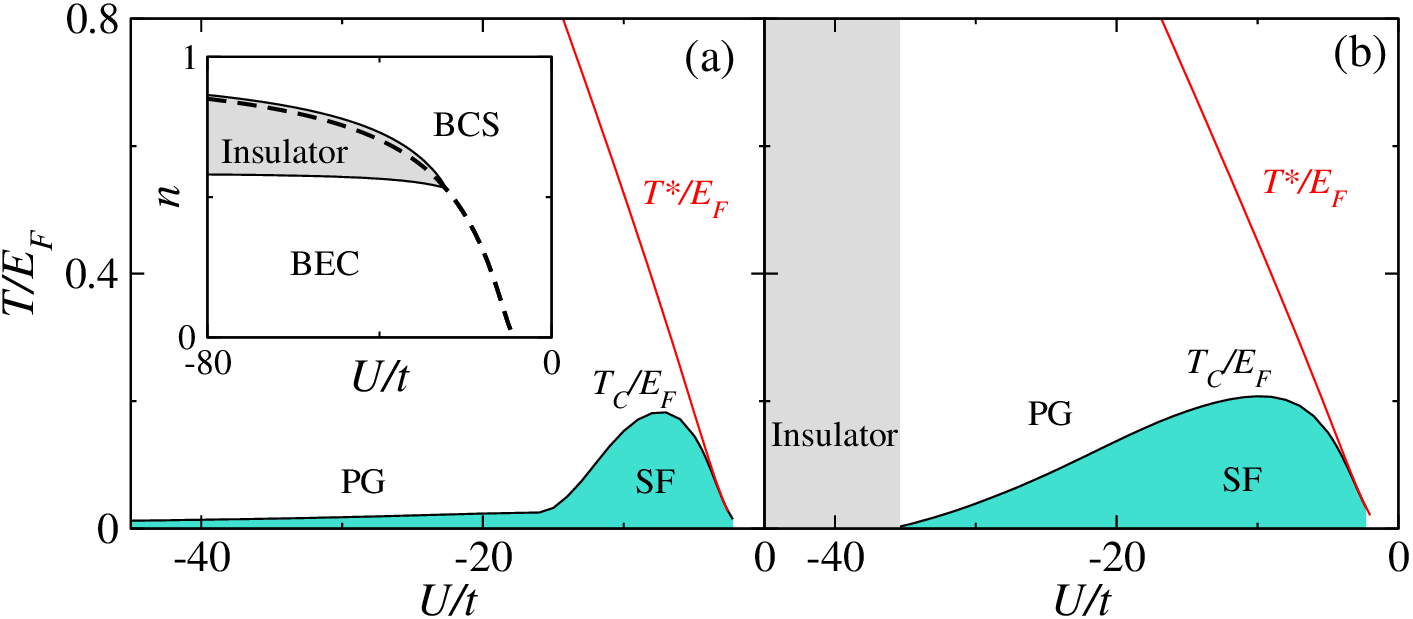}}
\caption{\label{fig:L1phase}Phase diagrams of one-channel attractive
  Hubbard model at (a) $n=0.3$ and (b) $n=0.7$. 'SF' denotes superfluid,
  'PG' denotes pseudogap phase and 'Insulator' in (b) schematically
  indicates breakdown of ground state superfluidity. Inset: $T=0$ phase
  diagram. Here the boundary separating BCS and BEC regimes is
  determined when $\mu$ reaches the bottom of the band. The gray shaded
  area shows where superfluid does not exist, corresponding to the
  'Insulator' regime. }
\end{figure}

In Fig.  \ref{fig:L1phase} we show typical phase diagrams at low and
high densities.  In contrast to $T_c$, the pairing onset temperature
$T^{*}$ (as estimated from the strict mean-field solution for $T_c$),
increases monotonically with $|U|/t$.  The high density phase
diagram, in particular, bears some similarity to its counterpart in high
temperature superconductors \cite{Reviews}.  In both figures, as in high
$T_c$ superconductors, we see an anti-correlation between the behavior
of $T_c$ and $T^*$, associated with lattice \cite{MicnasRMP} as well as
pseudgogap effects \cite{Chen1}.  The complete suppression of
superfluidity in Fig.~\ref{fig:L1phase}(b) can be shown explicitly to
come from localization of pairs; that is, the effective pair mass $M^*
\rightarrow \infty$.

In the inset of Fig.~\ref{fig:L1phase} we show the $T=0$ phase diagram,
in which ``BCS'' (``BEC'') denote states with $\mu$ higher (lower) than
the bottom of the band. The shaded regime corresponds to where there is
a non-superfluid ground state. The lowest value of $(|U|/t)_c \approx
25$ occurs at $n=0.53$. \textit{This is comparable in magnitude to the
  critical value $|U|/t\approx 35$ of the Mott-superfluid transition of
  the 3D boson (repulsive) Hubbard model} with filling factor $1$
\cite{oldBloch}, although the physical origin of the two insulators is
different.

In actual experiments, attractive interactions between the atoms are
generated by Feshbach resonance effects, which in principle, require a
two-channel description.  Previous work on the two-channel model in a
lattice concentrated on the superfluid-Mott transition in the
strongly-interacting regime near $n=2$ \cite{Carr05,Zhou05}. Other
recent work discussed band insulators in the weakly-interacting regime
\cite{Holattice}.  The generalization of our $T$-matrix formalism to the
two-channel model for Fermi gases was presented in \cite{JS2}.  To take
into account the closed-channel molecules, the Hamiltonian is extended
to $H=H_{f}+H_{b}+H_{fb}$, where $H_{b}=-\sum_{\mathbf{q}}
E_{\mathbf{q}}^{m} b^{\dagger}_{\mathbf{q}} b_{\mathbf{q}}$ describes
the hopping of molecules and $H_{fb}=\sum_{\mathbf{q,k}}
g(b^{\dagger}_{\mathbf{q}} c_{\mathbf{q}/2-\mathbf{k}\downarrow}
c_{\mathbf{q}/2+\mathbf{k} \uparrow}+\mbox{h.c.})$ describes conversion
between molecules and fermion pairs. Here $E_{\mathbf{q}}^{m} =
\epsilon_{\mathbf{q}}^{m}-2\mu+\nu$, $\nu$ is the magnetic detuning, and
$g$ is the coupling constant for molecule-pair conversion.  The strength
of the attractive pairing interaction is modified relative to the one
band case \cite{JS2} to
\begin{equation}
U_{eff}(Q)=U+g^{2}D_{0}(Q)
\end{equation}
where $D_{0}(Q)=1/(i\Omega_{n}-E_{\mathbf{q}}^{m})$ is the propagator
for non-interacting closed channel molecules. 
%and we define
The gap equation involves only $U_{eff}\equiv U_{eff}(Q=0)$.  The
density in the lattice now becomes $n=n_{f}+2n_{b}+2n_{b}^{0}$ where the
open-channel contribution $n_{f}$ is given by Eq.~(\ref{eq:1neq}) and
the closed-channel contribution comes from the molecular condensate
($2n_{b}^{0}$), and non-condensed molecules ($2n_{b}$).  The energy
dispersion of molecules has a Bloch-like dispersion, just as found for
the fermions. In the long wavelength limit, we may expand this Bloch
band dispersion as $\epsilon_{\mathbf{q}}^{m} \approx q^{2}/2M_{b}$,
where $M_{b}=2/t$ is the effective bare mass of the molecules. The
fermion pairs have a similar $q^2$ dispersion and, in general, hybridize
strongly with these closed-channel molecules.

In our Hamiltonian we have dropped a direct closed-channel boson-boson
repulsion. This was included in previous work \cite{Carr05,Zhou05} which
aimed to create Mott insulating phases associated with the closed
channel.  At $n=2$, because the closed channel represents a band which
is never completely filled (due to the presence of the open channel), we
find that localized states in the strict Mott sense are not obtainable
as is consistent with Ref.~\cite{Zhou05}.  In this way we argue that it
is appropriate here to drop the intra-closed-channel interactions.  We
note parenthetically that to obtain a Mott insulating state one
possibility is to treat the closed channel as composite fermion pairs
(or hard-core bosons \cite{MicnasRMP}) and to consider filling
\textit{above} $n=2$.

\begin{figure} 
\centerline{\includegraphics[clip,width=3.4in]{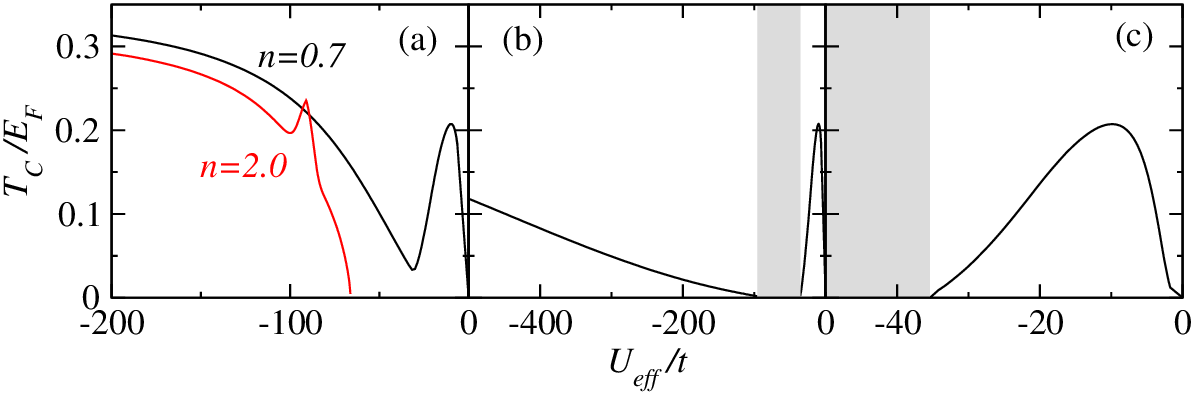}}
\caption{\label{fig:L2Tc} $T_c/E_F$ as a function of pairing interaction
  $U_{eff}/t$ in the two-channel model, as tuned by (a,b) detuning $\nu$
  and (c) $U$ for relatively (a,c) narrow and (b) broad Feshbach
  resonances. There parameters are: (a) $g/t=-60$, $U/t=-6$, $n=0.7$
  (black) and $2$ (red curve); (b) $n=0.7$, $g/t=-600$, $U/t=-6$; (c)
  $n=0.7$, $g/t=-60$, $\nu/t=500$.  The shaded regimes in (b) and (c)
  schematically indicate the new, non-integer-filling insulating
  states.}
\end{figure}

Figure \ref{fig:L2Tc} shows $T_c/E_{F}$ in the two-channel model as a
function of $U_{eff}/t$ for [(a) and (c)] narrow and (b) broad Feshbach
resonances. Here $U_{eff}$ is tuned via $\nu$ with fixed $U$ and $g$ in
(a) and (b) and via $U$ with fixed $\nu$ and $g$ in (c).  Figure
\ref{fig:L2Tc}(a) corresponds to the case in which there is a
considerable admixture of both closed and open channels. We take
$U/t=-6$ and consider a relatively narrow resonance, $g/t=-60$, as well
as two values of $n=0.7$ and $n=2$. For $n=0.7$, $T_c$ is first
suppressed by the opening of the pseudogap as $-U_{eff}$ increases; then
$T_c$ eventually increases with increasing $-U_{eff}$ as fermions are
converted into closed-channel molecules which have a finite BEC
transition temperature. For $n=2$, the cusp in the $T_c$ curve comes
from the van Hove singularity at $n_{f}=1$.  For a large range of
$U_{eff}$, $T_c$ for $n=2$ is effectively zero \cite{bandinsulator},
until a sufficient number of open channel pairs are converted to
closed-channel molecules.  Importantly, \textit{for this case, the
  insulating state thereby observed is a band insulator}.
%  This is consistent with
%earlier work \cite{Holattice} based on $T=0$.

When we decrease the participation of closed channel molecules by either
increasing $-U/t$ or $|g|/t$, insulating states (of the new form, at
non-integer filling) start to emerge.  In Fig.~\ref{fig:L2Tc}(b) we
demonstrate these insulating states by plotting $T_c/E_{F}$ for $n=0.7$
at $U/t=-6$ and a large $|g|/t=600$ (wide resonance) as a function of
$U_{eff}/t$ by tuning $\nu$. In Fig.~\ref{fig:L2Tc}(c) we plot the
counterpart curve for a narrower resonance ($g/t=-60$) as a function of
$U_{eff}/t$ by tuning $U$ at fixed high detuning $\nu$. In contrast to
the result shown in Fig.~\ref{fig:L2Tc}(a), by decreasing the fraction
of closed channel states, as in panels (b) and (c), we find localized
insulating phases.  This localization can be demonstrated by a
divergence in the effective pair mass.  In summary, insulating states at
non-integer filling are robust even in the presence of closed-channel
molecules.

The experimental implications of our work are readily testable, since,
fortunately, using Feshbach resonances, current cold atom experiments
\cite{Ketterlelattice} are able to simulate these attractive Hubbard
Hamiltonians.  Our principal result is the theoretical observation of
superfluid-insulator transitions in the ground state of the attractive
Hubbard model away from integer filling. To observe these new phases
experimentally one needs to consider average densities $n \approx 1$ and
sufficiently large $|U|/t$ of the order of, or larger than, that
required for superfluid-Mott insulator transitions in Bose gases
\cite{oldBloch}.  Because experimentally there is an additional
background harmonic trapping potential, $n$ can never be precisely
specified throughout the lattice \cite{Kohl05a} and thus it should be
possible to find extended regions with non-integer filling factors.  The
tunable parameters in optical lattice experiments are scattering length
and lattice potential depth $V_{0}$.  While it is relatively easy to
express $t$ in terms of $V_0$, the conversion of the on-site attraction
$U$ in terms of the scattering length and $V_0$ near unitarity is not as
straightforward as that in free space \cite{oldBloch}.

It is clear that the crucial test of the new phase diagram in Fig.
\ref{fig:L1phase}(b) does \textit{not} lie in distinguishing whether
$T_c$ is small or strictly zero. Rather with $T_c =0$ one can invoke
entropic considerations and deduce that signatures of this new ground
state will involve detecting some new form of (bosonic) order.  This may
be the analogue of a (pair density wave) phase which appears to be
present in high temperature superconductors, in the underdoped side of
the phase diagram \cite{Zhang2004}.

This work is supported by NSF Grant No.~PHY-0555325 and NSF-MRSEC Grant
No.~DMR-0213745. We thank Jit Kee Chin for helpful conversations.

\vspace*{-1ex} \bibliographystyle{apsrev}

%\bibliography{Review2}

\begin{thebibliography}{22}
\bibitem[{ban()}]{bandinsulator}
\bibinfo{note}{In the 2 channel limit as distinguished form the 1 channel case
  we find that $T_c$ at $n=2$ is never strictly zero but the deviations are
  negligible.}
\end{thebibliography}

\end{document}